\pdfoutput=1
\documentclass[a4paper,american,citeautoscript,floatfix,pdftex,showpacs,superscriptaddress,%
aps,pra,%
longbibliography,%
reprint,twocolumn,
]{revtex4-1}
\usepackage{amsfonts,amsmath,amssymb}
\usepackage{graphicx}
\usepackage[T1]{fontenc}
\usepackage[utf8]{inputenc}
\usepackage{microtype}
\usepackage{textcomp}
\usepackage[textsize=footnotesize]{todonotes}
\usepackage{hyperref}
\usepackage{hypernat}
\usepackage[todos]{CMImacros}

\graphicspath{{./figures/}} 
\newcommand{\cfeldesy}{\affiliation{Center for Free-Electron Laser Science, Deutsches
      Elektronen-Synchrotron DESY, Notkestrasse 85, 22607 Hamburg, Germany}}%
\newcommand{\uhhcui}{\affiliation{The Hamburg Center for Ultrafast Imaging, University of Hamburg,
      Luruper Chaussee 149, 22761 Hamburg, Germany}}%
\newcommand{\uhhphys}{\affiliation{Department of Physics, University of Hamburg, Luruper Chaussee
      149, 22761 Hamburg, Germany}}%
\newcommand{\nijmegen}{\affiliation{Institute for Molecules and Materials, Radboud Universiteit,
      Houtlaan 4, 6525 XZ Nijmegen, The Netherlands}}%
\newcommand{\jkemail}{\email[Corresponding Author. Email:~]{jochen.kuepper@cfel.de}}%
\newcommand{\cmiweb}{\homepage[URL:~]{https://www.controlled-molecule-imaging.org}}%

\makeatletter%
\renewcommand*{\@fnsymbol}[1]{\ensuremath{\ifcase#1\or \|\or *\or **\or \mathparagraph\or
      \mathsection\or \dagger\or \ddagger\or \dagger\dagger \or \ddagger\ddagger \else\@ctrerr\fi}}
\makeatother

\begin{document}
\title{Light-sheet imaging for the recording of transverse absolute density distributions \\ of
   gas-phase particle-beams from nanoparticle injectors}%
\author{Lena Worbs}\cfeldesy\uhhphys%
\author{Jannik Lübke}\cfeldesy\uhhphys%
\author{Nils Roth}\cfeldesy\uhhphys%
\author{Amit K.\ Samanta}\cfeldesy%
\author{Daniel A.\ Horke}\cfeldesy\nijmegen\uhhcui%
\author{Jochen Küpper}\jkemail\cmiweb\cfeldesy\uhhphys\uhhcui%
\date{\today}%
\begin{abstract}\noindent%
   Imaging biological molecules in the gas-phase requires novel sample delivery methods, which
   generally have to be characterized and optimized to produce high-density particle beams. A
   non-destructive characterization method of the transverse particle beam profile is presented. It
   enables the characterization of the particle beam in parallel to the collection of, for instance,
   x-ray-diffraction patterns. As a rather simple experimental method, it requires the generation of
   a small laser-light sheet using a cylindrical telescope and a microscope. The working principle
   of this technique was demonstrated for the characterization of the fluid-dynamic-focusing
   behavior of 220~nm polystyrene beads as prototypical nanoparticles. The particle flux was
   determined and the velocity distribution was calibrated using Mie-scattering calculations.
\end{abstract}
\maketitle

\section{Introduction}%
Knowledge of the structure of biological molecules, such as proteins or viruses, is fundamental for
understanding their function. A recently pioneered approach for directly recording high-resolution
structures of intact single molecules is single-particle coherent diffractive imaging using x-ray
free-electrons lasers (XFELs)~\cite{Neutze:Nature406:752, Barty:ARPC64:415}. To reconstruct a
three-dimensional molecular structure, this approach requires the collection of a large number of
individual diffraction patterns from single molecules~\cite{Bogan:AST44:i, Ekeberg:PRL114:098102}.
In order to achieve this within the limited amount of time available at central XFEL facilities,
therefore, requires a high particle flux in the gas-phase. Furthermore, as the samples investigated
get smaller in size, now approaching the limit of single proteins, the necessary x-ray intensity for
recording a single-shot diffraction pattern increases. Experimentally, this higher intensity is
typically achieved by focusing the x-ray beam to a smaller spot, in the most extreme cases to sizes
of only $\ordsim100$~nm. This places stringent demands on the employed sample delivery methods,
typically aerodynamic lens stacks (ALS)~\cite{Liu:AST22:293, Hantke:NatPhoton8:943, Roth:JAS124:17},
and requires their characterization and optimization prior to XFEL experiments with laboratory-based
methods.

Any characterization method for nanoparticle injectors would ideally reconstruct the full
six-dimensional phase space of nanoparticles emitted, and would do so on-the-fly, \emph{in situ},
non-destructive, and universally for any nanoparticle. Furthermore, the simultaneous
characterization of sheath gas flows would be advantageous, but that seems to be well delegated to
offline analysis~\cite{Horke:JAP121:123106}. None of the currently available nanoparticle-imaging
methods fulfills all these requirements~\cite{Awel:OE24:6507}. The most commonly employed method is
optical imaging of the particle stream using side-illumination with a laser
beam~\cite{Awel:OE24:6507}. This allows one to measure the longitudinal position of nanoparticles,
as well as their longitudinal velocity through the use of double-pulse lasers~\cite{Hantke:ec5009},
commonly termed \emph{particle imaging velocimetry}, or by recording light streaks from the
particles through the use of appropriate illumination or exposure times~\cite{Awel:OE24:6507}. While
this side-view light-scattering approach fulfills the requirements for \emph{in situ} operation and
largely the universality for any nanoparticle, it does not detect asymmetries in the transverse
nanoparticle beam profile without scanning the laser focus through the particle beam. For larger
particle beams this approach can, furthermore, suffer from a mismatch between particle beam size and
focal depth of the used imaging objective.

A direct way to record the transverse profile of particles has so far only been available through
so-called dusting, where particles are caught on a sticky surface in-vacuum, which is subsequently
imaged~\cite{Awel:OE24:6507}. This approach has the obvious drawback of being destructive.
Furthermore, it does not allow the recording of absolute particle densities, as the sticking
probability of particles is not unity, nor of particle velocities.

Here, we present a different approach, combing the advantages of laser-based scattering microscopy
with the ability to measured transverse particle positions: light-sheet imaging (LSI). A sheet of
light is generated and the scattered light from the particles passing through the sheet is collected
using a microscope. This method provides a simple, non-destructive, and \emph{in situ} method to
record transverse particle beam profiles, including absolute number densities, as well as the
velocity of particles passing through the light sheet.

\section{Experimental Setup}
\begin{figure}[t]
   \includegraphics[width=\linewidth]{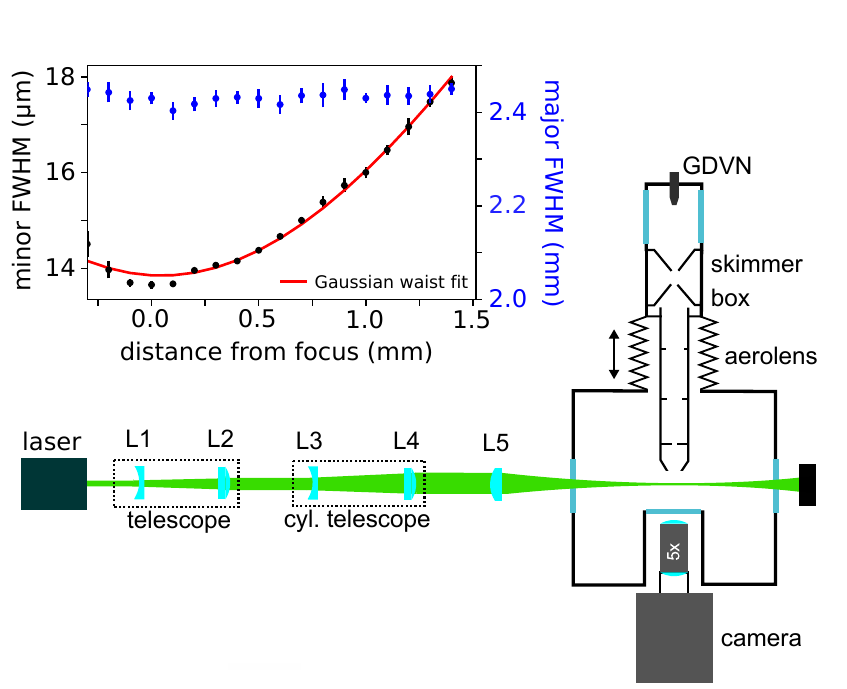}
   \caption{Schematic of the light-sheet imaging setup for the characterization of nanoparticle
      beams emerging from an aerodynamic lens stack. The light sheet is generated using a spherical
      telescope, a cylindric-lens telescope, and a spherical focusing lens to create an elliptical
      focus. The inset shows the measured values of the minor and major widths along the light beam
      and the corresponding Gaussian beam-waist fit to determine the Rayleigh length $z_R=1.5$~mm.}
   \label{fig:setup}
\end{figure}

A schematic overview of the LSI setup is shown in \autoref{fig:setup}. It mainly consisted of three
parts: the optical setup for the generation of the light sheet from a continuous-wave laser, an ALS
to generate a nanoparticle beam, and a microscope-detector setup for recording the scattered light.

The light sheet was generated from a continuous-wave laser system (Coherent Verdi V, 5~W, 532~nm),
operated at 0.5~W and with vertical polarization. The initial beam diameter of $2.25(23)$~mm is
increased by a factor of two using a Galilean telescope ($f_{L1}=-50$~mm, $f_{L2}=100$~mm);
throughout the manuscript all laser-beam widths are specified to the $1/e^2$ intensity. A second
cylindrical-lens telescope ($f_{L3}=-50$~mm, $f_{L4}=150$~mm) further increases the vertical
diameter of the beam by a factor of three, yielding an elliptical beam with a vertical width of
$h=13.5$~mm and a horizontal width of $w=4.1$~mm. This beam is focused using a cylindrical lens
($f_{L5}=300$~mm), generating the light sheet in the center of the vacuum chamber. The created light
sheet has a thickness of 24.8~\um (14.6~\um full-width-half-maximum, FWHM), a Rayleigh length of
$z_R=1.5$~mm, and a horizontal width of $w=4.1$~mm. The intensity in the focus is
$\ordsim2.8\times10^3~\Wpcmcm$, well below the typical damage threshold for nanoparticles.

The nanoparticle beam was generated using a previously described ALS~\cite{Roth:JAS124:17}. Briefly,
220~nm polystyrene beads in solution (Alfa Aesar, particle size $220\pm17.3$~nm) were diluted to
$5\times10^6$~particles/ml and aerosolized using a gas-dynamic virtual nozzle
(GDVN)~\cite{DePonte:JPD41:195505, Beyerlein:RSI86:125104}. Following a differential pumping stage,
nanoparticles entered the ALS, which produced a tightly collimated and focused particle stream in
the vacuum chamber; the chamber pressure was typically held at $6\times10^{-5}$~mbar during
experiments. The entire ALS system was placed on a motorized $xyz$-translation stage to allow
accurate positioning of the particle stream.

Particles emerging from the the ALS passed through the light sheet and the scattered light was
collected by a camera-based microscope system. This consisted of a long working-distance objective
(Edmund Optics, $5\times$ magnification, numerical aperture 0.14, working distance 34~mm, depth of
field 14~\um) and a high-efficiency sCMOS camera (Photometrics Prime 95B, quantum efficiency 0.95 at
532~nm, $1200\times1200$~pixels). This yields a nominal resolution of 0.54~px/\um. Images were
collected with a 1~ms exposure time and at a frame rate of 82~fps at full-frame size. Collected
images were analyzed using a centroiding algorithm based on Hessian
blob-finding~\cite{Marsh:SciRep8:978}, yielding sub-pixel particle positions as well as an accurate
estimate of the number of recorded photons per particle.

The data shown were typically recorded for 2\,000 frames and limited to a region of interest
covering $350\times350$~pixels. Transverse beam profiles were generated as 2D histograms of the
recorded particle positions. For the full three-dimensional reconstruction of the particle beam, the
distance between the ALS exit and the light sheet was varied to build up a series of 2D transverse
profiles. To recover the velocity of particles passing through the sheet, the recorded scattering
intensity was compared with calculations based on Mie theory, performed using a homebuilt Python
script based on the freely available Bohren and Huffman code~\cite{Bohren:ScatteringLight}, taking
into account the experimental parameters, \eg, the scattering angle, the numerical aperture, and the
particle size.

\section{Results and Discussion}
An example image showing two nanoparticles passing through the light sheet simultaneously is shown
in \autoref[a]{fig:results1}. Here, the concentration of the nanoparticle solution was reduced from
stock to ensure that camera frames contained no more than 2 particles.
\begin{figure}
   \includegraphics[width=\linewidth]{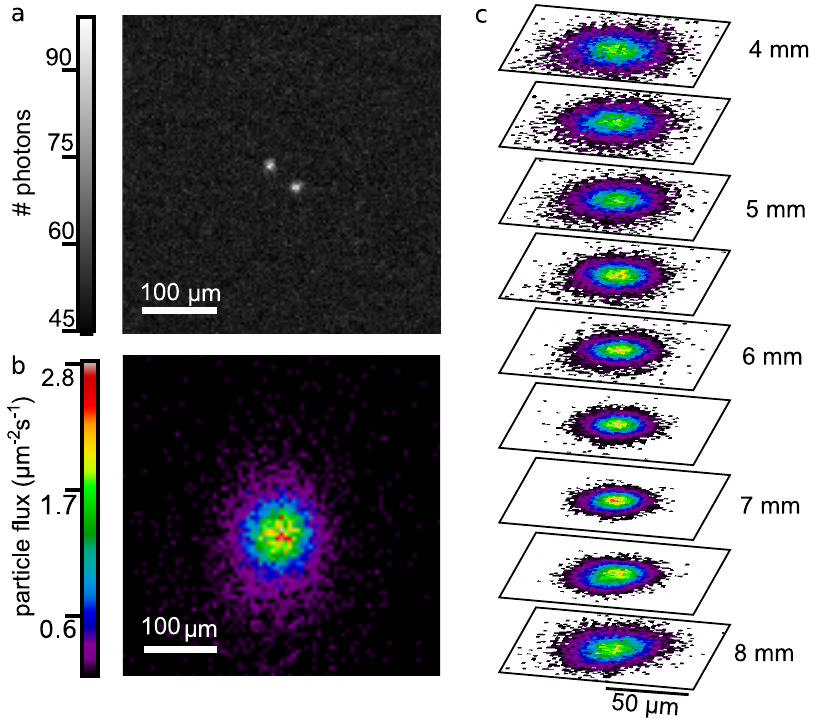}
   \caption{(a) Raw image of a single camera frame containing two particles. (b) Transverse beam
      profile 6~mm below the injector tip, shown as a 2D histogram of the determined particle
      positions; the colorscale represents particle flux. (c) 3D beam profile shown as individual 2D
      histograms measured at different distances from the ALS exit. The colorscale is the same as in
      (b).}
   \label{fig:results1}
\end{figure}
An example of a measured full particle beam profile is shown in \autoref[b]{fig:results1}, recorded
6~mm below the ALS injector. In \autoref[c]{fig:results1}, the 3D beam profile is shown as
individual 2D histograms recorded at several distances. These measurements directly reveal, for
instance, any astigmatism that might be present in the particle beam. From \autoref[c]{fig:results1}
it is clear that moving away from the injector tip the particle-beam widths changes, showing the
characteristic focusing behavior of the injector~\cite{Roth:JAS124:17}. Recorded particle beam
profiles typically show a Gaussian intensity distribution and the evolution of the particle beam can
be accurately described by fitting a 2D Gaussian function to the recorded data, thus allowing for an
easy evaluation of the beam widths and asymmetry through the minor and major axis of the fitted 2D
Gaussian. The analysis of the size at 7~mm downstream the injector reveals that, without any
specific optimization of the ALS, we generated a particle beam with a focus FWHM of $44.6(33)~\um$.

Since the measurement is based on counting individual particles, the histograms reveal the absolute
particle flux, that is, the number of particles per area per second. The rate of particles is
recovered from the observation time, which due to the continuous illumination is simply given by the
number of frames multiplied by their exposure times. The particle flux can then be calculated from
the number of particles detected per pixel or histogram bin. The peak particle flux per $\um^2$ per
second as a function of the distance from the ALS exit is shown in \autoref[a]{fig:results2}. Under
the given conditions, the maximum particle flux of $4.3~\text{particles}/\um^2/\text{s}$ is measured
7.0~mm below the injector tip.
\begin{figure}
   \includegraphics[width=\linewidth]{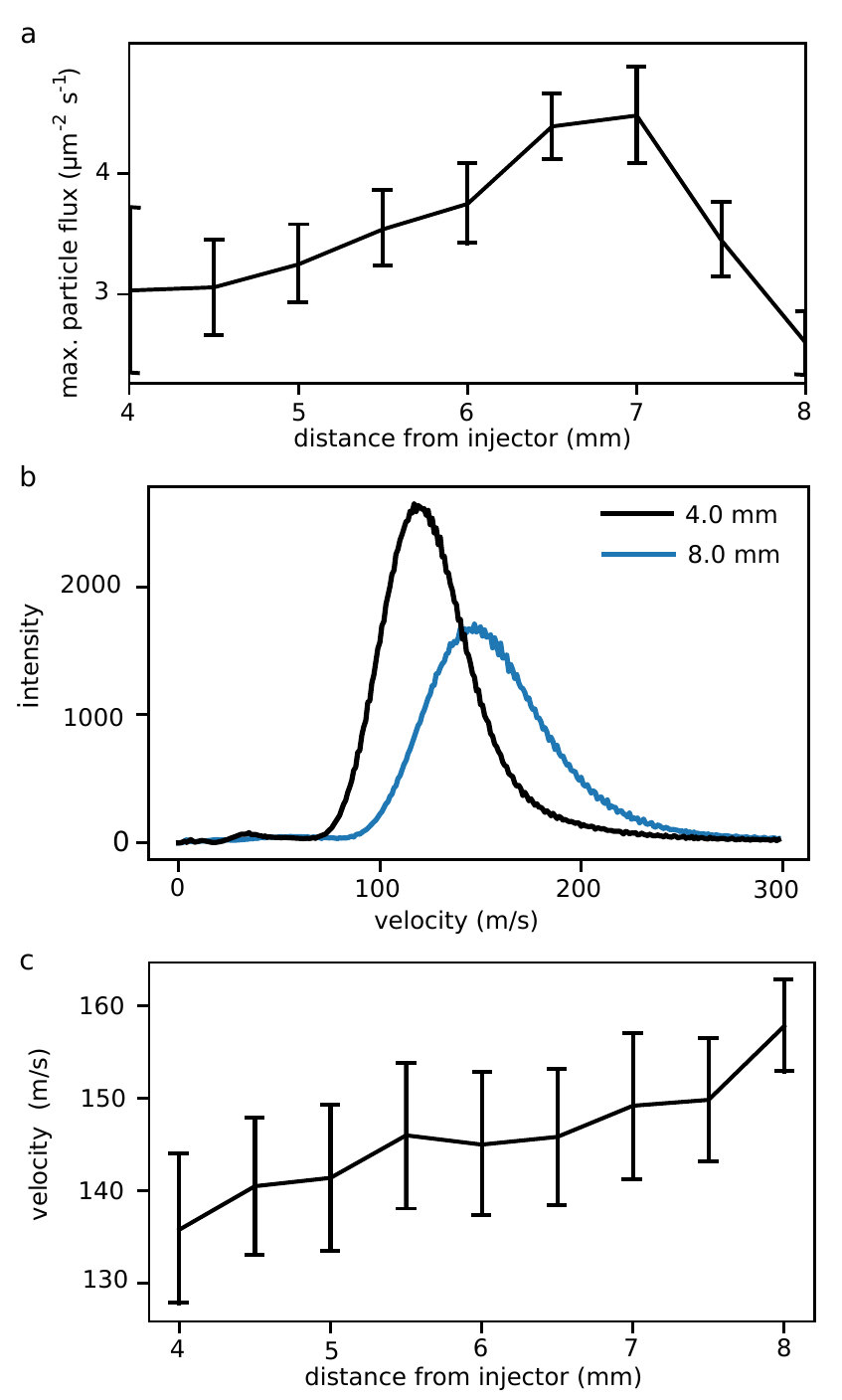}
   \caption{(a) Maximum particle flux as a function of the distance from the ALS, vertical bars
      represent the standard errors. (b) Velocity distributions measured at distances of 4~mm
      (black) and 8~mm (blue) from the ALS. (c) Mean of the velocity distribution depending on the
      distance from the injector.}
   \label{fig:results2}
\end{figure}

In order to recover the velocity of particles passing through the light sheet, the recorded
intensity distribution from single particles was modeled as Mie scattering. For each recorded
particle we calculated its velocity, taking into account that the particle size is given by the
manufacturers size distribution of $220\pm17$~nm. This results in a velocity probability
distribution for every single measured particle. The velocity distribution of all particles measured
at a certain distance from the ALS exit is the sum of the individual velocity probability
distributions, see \autoref[b]{fig:results2}. At 4~mm (black curve) downstream the ALS exit, the
velocity distribution is centered around 122~m/s and the FWHM is 49~m/s. At 8~mm (blue curve) the
mean velocity is 151~m/s with a FWHM of 66~m/s. \autoref[c]{fig:results2} shows the mean velocity of
the particle stream as a function of distance to the ALS exit. Both, mean velocity and the width of
the velocity distribution increased with distance from the ALS exit, consistent with recent
observations~\cite{Hantke:ec5009}. We ascribe the increasing velocity to acceleration by the helium
gas co-emerging from the ALS, hinting at the need to correlate this acceleration with measurements
of the gas density~\cite{Horke:JAP121:123106} in future work. The observed velocities above 100~m/s
show that the particles are fast enough to clear the interaction region with an x-ray beam in a
typical single-particle imaging experiment in between two pulses, assuming a \um beam size,
including tails, and the 4.4~MHz repetition rate of the EuXFEL~\cite{Bean:JO18:074011}.

\section{Conclusion} Light-sheet imaging (LSI) provides a non-destructive measurement technique for
transverse nanoparticle beam profiles. It is general and applicable to beams from any gas-phase
nanoparticle-beam injector. Here, we characterized a nanoparticle beam generated from an aerodynamic
lens stack. A light sheet with a width comparable to the depth of view of the used microscope
objective was generated using cylindric lenses. The scattered light from the nanoparticles was
collected and utilized to determine the particle positions in order to measure the transverse beam
profile and the absolute particle flux. The number of collected scattered photons was used for
determining the velocity of the nanoparticles.

In current single-particle diffractive-imaging experiments, the initial particle
concentration in solution is in the order of 10$^{14}$ particles/ml, which increases the maximum
particle flux in our experiment to $\ordsim10^{7}~\text{particles}/\um^2/\text{s}$, corresponding to
$\ordsim10^{4}~\text{particles}/\um^2/\text{frame}$ at 1~ms exposure time. In this case, an \emph{in
   situ} characterization would be enabled through limited effective illumination, \eg, using a
pulsed illumination scheme.

Future single-particle diffractive-imaging experiments on biological molecules, with particle sizes
of $\ordsim10$~nm, require the benchmarking of ALSs for such small particle sizes. The use of
light-sheet imaging for such small particles requires increased laser powers. In our specific setup,
\eg, with a 5~W green laser, the detection of particles is limited to particles down to
$\ordsim100$~nm to be distinguishable from noise. The detection of smaller particles, \eg, 10~nm
proteins with typical velocities from an ALS of 120~m/s, a cw laser power of 54~MW would be
necessary, which is practically infeasible~\cite{Deppe:OE23:28491}. However, pulsed illumination
mimicking the XFEL temporal pulse profiles would enable equivalent information and still be feasible
with standard laser technology routinely available at XFEL facilities.

\begin{acknowledgments}\noindent%
   This work has been supported by the European Research Council under the European Union's Seventh
   Framework Programme (FP7/2007-2013) through the Consolidator Grant ``COMOTION''
   (ERC-614507-Küpper), by the Helmholtz Gemeinschaft through the ``Impuls- und Vernetzungsfond'',
   and by the Deutsche Forschungsgemeinschaft (DFG) through the clusters of excellence ``Center for
   Ultrafast Imaging'' (CUI, EXC~1074, ID~194651731) and ``Advanced Imaging of Matter'' (AIM,
   EXC~2056, ID~390715994).
\end{acknowledgments}

\bibliography{string,cmi}
\end{document}